\documentclass[apjl]{emulateapj}

\usepackage{graphicx}

\begin{document}

\shorttitle{Metallicity and LMXBs}
\shortauthors{Ivanova et al.}

\bibliographystyle{apj}

\title{On the origin of the metallicity dependence in dynamically formed extragalactic low-mass X-ray binaries}

\author{N.\ Ivanova$^{1,2,3}$, T.\ Fragos$^4$, D.-W.\ Kim$^4$, G.\ Fabbiano $^4$, J.L.\ Avendano Nandez$^1$,  J.C.\ Lombardi$^5$, G.\ R.\ Sivakoff$^1$, R.\ Voss $^6$, A.\ Jord\'an $^{7,8}$
}
\altaffiltext{1}{Department of Physics, University of Alberta, Edmonton, AB T6G 2E1, Canada}
\altaffiltext{2}{Canada Research Chair in Astrophysics}
\altaffiltext{3}{nata.ivanova@ualberta.ca}
\altaffiltext{4}{Harvard-Smithsonian Center for Astrophysics, 60 Garden Street, Cambridge, MA 02138, USA}
\altaffiltext{5}{Department of Physics, Allegheny College, Meadville, PA 16335, USA}
\altaffiltext{6}{Department of Astrophysics/IMAPP, Radboud University Nijmegen, PO Box 9010, 6500 GL, Nijmegen, The Netherlands}
\altaffiltext{7}{Departamento de Astronom{\'i}a y Astrof{'}{\'i}sica, Pontificia Universidad Cat{\'o}lica de Chile, 7820436 Macul, Santiago, Chile}
\altaffiltext{8}{The Milky Way Millennium Nucleus, Av. Vicu\~{n}a Mackenna
       4860, 7820436 Macul, Santiago, Chile}

\begin{abstract}
Globular clusters (GCs) effectively produce dynamically-formed low-mass X-ray binaries (LMXBs). 
Observers detect $\sim100$ times more LMXBs per stellar mass in GCs compared to stars in the fields of galaxies. 
It has also been observationally established that metal-rich GCs are about 3 times more likely to contain an X-ray source than their metal-poor counterparts.
Recent observations have shown that this ratio holds in extragalactic GCs for all bright X-ray sources with $L_{\rm X}$ between $2\times 10^{37}$ and $ 5\times 10^{38}\,\rm erg\, s^{-1}$. 
In this {\it Letter}, we propose that the observed metallicity dependence of LMXBs in extragalactic GCs
can be explained by the differences in the number densities and average masses of red giants in populations of different metallicities.
Red giants serve as seeds for the dynamical production of bright LMXBs via two channels -- binary exchanges and physical collisions -- and 
the increase of the number densities and masses of red giants boost LMXB production, leading to the observed difference.
We also discuss a possible effect of the age difference in stellar populations of different metallicities.
\end{abstract}

\section{Introduction}

The preference of bright low-mass X-ray binaries (LMXBs), with $L_{\rm X}>10^{36}\,\rm erg\, s^{-1}$,
to reside in metal-rich globular clusters (GCs) was first noted for our Galaxy and M31 
\citep{1993ASPC...48..156G, 1995ApJ...439..687B}.
Extragalactic observations with a much larger sample of GCs and observed bright 
LMXBs, but also at a higher cut-off luminosity, $L_{\rm X}>10^{37}\,\rm erg\, s^{-1}$, have confirmed this tendency 
\citep[e.g.,][]{2002ApJ...574L...5K,2004ApJ...606..430M, 2004ApJ...613..279J,2006ApJ...647..276K,2007ApJ...662..525K,2007ApJ...660.1246S}.
Similar, albeit weaker statistically, trends have also been seen in the GCs of M31 for LMXBs with 
$L_{\rm X} \gtrsim 10^{36}\,\rm erg\, s^{-1}$ \citep{2004ApJ...616..821T,2010MNRAS.407.2611P}.
As the formation rate of bright LMXBs in GCs exceeds
that of the field by a factor of about 100, it is commonly accepted that bright LMXBs in GCs 
are formed dynamically (the idea was first proposed by \citealt{1975ApJ...199L.143C}, for an overview see \citealt{2006csxs.book..341V}).
The origin of the preference of bright LMXBs for metal-rich  clusters could lie in 
pre-dynamical enhancements of the constituents that eventually form into LMXBs, 
enhancements during dynamical LMXB formation, or both.
A recent set of deep observations of extragalactic LMXBs 
has strengthened the statement that metal-rich clusters 
are $\sim3$ times more likely 
to have bright LMXBs \citep{Kim12}, clarifying that 
this ratio holds across the range of X-ray luminosities 
from $2\times 10^{37}\,\rm erg\, s^{-1}$ to $5\times 10^{38}\,\rm erg\, s^{-1}$.
We show here that these latter observations provides the 
first statistically significant constraint on a type of a donor 
in the observed extragalactic bright GC-LMXB population.

\section{Dynamical formation of LMXBs in GCs}

There  are three  sub-populations of  LMXBs, depending on the evolutionary state of the donor star, 
that can potentially contribute to the X-ray luminosity function (XLF) of a dynamically formed X-ray binary 
(XRB) population in a GC: LMXBs with main sequence (MS) donors, LMXBs with subgiant or giant donors 
(hereafter, we refer to both giant and sub-giant low-mass stars as red giants, RGs), 
and ultra-compact XRBs (UCXBs) with degenerate, white dwarf (WD) donors. 

 The vast majority of XRBs in GCs should be accreting onto neutron stars (NSs) \citep[formation channels of XRBs with a black hole (BH) accretor differ 
from those with a NS accretor, see ][]{2004ApJ...601L.171K, 2010ApJ...717..948I}, which predominantly formed through either electron-capture supernovae or accretion-induced collapse \citep{2008MNRAS.386..553I}; the low escape velocity in GCs leads to the loss of almost all NSs formed via normal core-collapse with a large native kick. This results in a lighter population of NSs in GCs, with initial masses of $M_{\rm NS}\sim 1.28\pm0.06\,\rm M_{\odot}$ \citep{1996ApJ...457..834T}.

LMXBs in dense stellar systems are most efficiently formed from binaries with a NS and a MS donor \citep{2008MNRAS.386..553I}.
The strongest formation channels among MS donors are binary exchange interactions with a NS or the creation of a NS-MS binary  
following the accretion induced collapse of a WD in an already dynamically formed WD-MS binary. Tidal capture is significantly less efficient.
Although the formation rate of NS-MS systems is estimated to be the highest among the three sub-populations of donors \citep{2008MNRAS.386..553I}, 
the mass-transfer (MT) rate that these systems with low-mass MS donors can drive  is very low. 
As a consequence, most NS-MS LMXBs are transient with low duty cycles and
low outburst luminosities \citep{2008ApJ...683..346F,2009ApJ...702L.143F,2011A&A...526A..94R}. 
As such, they are not expected to have any contribution to the observed XLF at the luminosity range 
of interest here ($L_{X}>2\times 10^{37}\,\rm erg\, s^{-1}$).
While persistent NS-MS LMXBs can be formed in only metal-rich GCs (for old populations systems) and still play 
a significant role for the metallicity dependence in our Galactic LMXBs \citep{2006ApJ...636..979I}, 
their persistent luminosities are also below the limiting luminosity to which extragalactic LMXBs are usually observed.
Note that the maximum luminosity of 4U 1746-37, the only bright LMXB in Galactic GCs that most likely has a MS donor, is also below this threshold \citep{2001A&A...368..451S}.

RGs play a significant role in the dynamical formation of NS LMXBs, influencing  two important formation channels.
First, RGs can physically collide with NSs and form compact binaries that consist of a WD and a NS; such binary systems may later 
become UCXBs  \citep{2005ApJ...621L.109I}.
Second, following the formation (via an exchange encounter) of an eccentric NS-MS binary, the MS donor may overfill its Roche lobe as it evolves to become a subgiant or giant. 
In both cases, these binaries will appear as persistent LMXBs with $L_{\rm X}> 2\times 10^{37}\,\rm erg\, s^{-1}$  for a 
fraction of their MT evolution \citep{2008ApJ...683..346F,2009ApJ...702L.143F}.

Dynamical formation of bright LMXBs is determined by the probabilities of two events: first, some seed binary with a NS has to be formed in a dynamical encounter (see \S3), and second, this binary must start MT that is fast enough to appear as a bright LMXB (see \S4).

\section{Enhancement  of the dynamical encounters rate}
\label{sec:encounter_rate}

The total number of physical collisions (PCs) between RGs and NSs is $N_{\rm PC}=n_{\rm RG} n_{\rm NS} \sigma_{\rm RG,NS} v_{\infty}$, where $n_{\rm RG}$ and $n_{\rm NS}$ are the number densities of  RGs  and  NSs accordingly, 
$v_\infty$ is the relative velocity at infinity and $\sigma_{\rm RG,NS}$ is the  cross-section of an encounter: 

\begin{equation}
\sigma_{\rm RG,NS}= \pi r_{\rm P}^2 \left (1+\frac{2G(M_{\rm RG}+M_{\rm NS})}{r_{\rm P} v_{\rm \infty}^2} \right ) \ .
\end{equation}
\noindent Here $r_{\rm P}$ is the initial closest approach during an encounter.
The number of binary exchange encounters, $N_{\rm BE}$, is calculated similarly, replacing $n_{\rm RG}$ with the number density of seed binaries and setting $r_{\rm P}\approx 2 a$, where $a$ is the orbital separation in a pre-exchange binary containing a RG or a MS star that will evolve into a RG.

\begin{figure}
\plotone{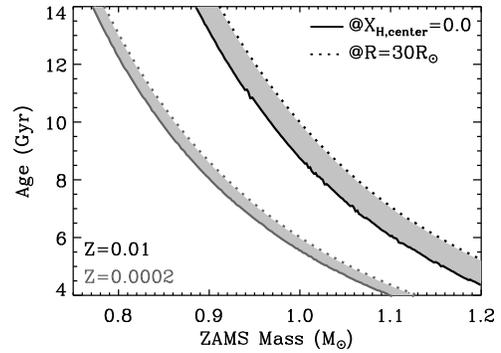}
\caption{ Age of star when it exhausts the hydrogen at its center (solid lines) and when it reaches a radius of $30\,\rm R_{\odot}$ (dotted lines) as a function of its zero age main-sequence mass, for two metallicity values, Z=0.01 (black lines) and Z=0.0002 (grey lines), typical of red and blue GC respectively.\label{Age_MS_RG}}
\end{figure}

Assuming all stars in a GC are coeval, all RGs in a GC have about the same mass, which only varies by a few percent 
compared to the MS turn-off mass of the specific GC.
This mass range is metallicity dependent since the life-times of metal-rich stars are longer than those 
of metal-poor stars, both for their MS life-time, $\tau_{\rm MS}$, and their life-time as a RG, $\tau_{\rm RG}$.
Figure~\ref{Age_MS_RG} shows the ages of a star when it exhausts the hydrogen at its center (end of MS) 
and when it reaches a radius of $30\,\rm R_{\odot}$ as a function of its zero age MS 
mass, for two metallicity values (Z=0.01 and Z=0.0002), typical of red  (metal-rich) and blue (metal-poor) GCs respectively.
The stellar evolution tracks used to create Figure~\ref{Age_MS_RG} were calculated using the {\tt MESA} stellar
evolution code \citep{2011ApJS..192....3P}. 

Although the choice of $30\,\rm R_{\odot}$ in Figure~\ref{Age_MS_RG} is not strictly rigid, it is justified by two reasons.
First, it is approximately the maximum radius that a RG can have as part of a binary system in a GC environment. 
The characteristic collision time for a binary is 

\begin{equation}
\tau_{\rm coll} \approx 0.8 {\rm Gyr} \frac{v_{10}}{n_5 a_{100} (M_{\rm bin} + \langle M \rangle )}
\end{equation} 
\noindent Here $n_5=n/(10^5 {\rm \, pc}^{-3})$, where $n$ is the stellar number density, 
$M_{\rm bin}$ is the total binary mass in $M_\odot$, $\langle M \rangle$ is the mass of an average single star in $M_\odot$, 
$v_{10} = v_\infty/(10 {\rm \, km\ s}^{-1}) $ and $a_{100}=a/(100 \, R_\odot)$ where $a$ is the binary separation.
In a typical dense cluster, a binary with $a \sim 100 \,R_\odot$ 
(i.e., about the size of the Roche lobe radius for a RG of $\sim 30 R_\odot$) 
will be perturbed by a collision during its 
$\tau_{\rm RG}$. This likely leads to an increased eccentricity that starts MT before the RG reaches $\sim 30 \, R_\odot$.
Larger orbital separations would correspond to very wide (soft) binaries that would be quickly disrupted via strong dynamical interactions.
Second, larger RGs are unlikely  to form a compact binary after a PC \citep{2005ApJ...621L.109I, 2006ApJ...640..441L}.
Since any RG evolution from $R_{\rm RG} \sim 30 R_\odot$ through the tip of a RG branch takes less than 1\% of $\tau_{\rm RG}$, the exact upper limit of $R_{\rm G}$ beyond $30 R_\odot$ that can make an LMXB/UCXB does not significantly change $\tau_{\rm RG}$.

Figure~\ref{Age_MS_RG} shows that the mass range at which a giant star can exist in a metal-rich GC is 
significantly larger than the one corresponding to a metal-poor GC of the same age. 
For a typical age of a Milky Way GC, $11\,\rm Gy$, a red cluster has RGs with masses 
          $M_{\rm RG, red}\simeq 0.946-0.972\,\rm M_{\odot}$, 
while a metal-poor GC  had  RGs with  masses 
          $M_{\rm  RG, blue}\simeq 0.826-0.838\,\rm M_{\odot}$.
Assuming a \cite{2001MNRAS.322..231K} initial mass function, this difference in mass ranges of RGs 
translates to a number density of giant stars $n_{\rm RG}$ that is $\sim 60\%$ higher in red clusters compared to blue clusters.
If extragalactic red GC have a younger stellar population \citep[see, e.g.][]{2010ApJ...708.1335W}, the difference in number density can be increased even further: e.g., RG in a metal-rich GCs of $\sim 8\,\rm Gyr$ will be about twice more abundant than in a $\sim 11\,\rm Gyr$ metal-poor GC.%
\footnote{The determination of an exact GC age is not possible due to a number of observational uncertainties \citep[for a review, see][]{2006ARA&A..44..193B} and theoretical uncertainties \citep[e.g., the presence of binaries also changes theoretical ages estimate;][]{2012arXiv1205.4310F}.} 
The total number of dynamical encounters leading to LMXBs formation in red GCs is $\sim 1.6$--$2$ higher compared to blue GCs due to the increase in $n_{\rm RG}$.

The average mass of the RGs in a cluster also produces a secondary enhancement through cross section of encounters, $\sigma_{\rm RG,NS}$. For PCs, the second term in the brackets of Eq.~(1) is much larger than one, leading to the mass-dependence increase in the cross-sections of encounters in red GCs compared to blue GCs by a factor of $(M_{\rm RG, red}+M_{\rm NS})/(M_{\rm RG, blue}+M_{\rm NS})$. For a typical GC age of $11\,\rm Gyr$, the average mass of a giant star in a metal-rich GC is higher ($M_{\rm RG, red}\sim 0.96\,\rm M_{\odot}$) compared to a metal-poor cluster ($M_{\rm RG, blue}\sim 0.83\,\rm M_{\odot}$). In a younger stellar population, $\sim 8\,\rm Gyr$, $M_{\rm RG, red} \sim1.05\,\rm M_{\odot}$. The increased cross section provides a mild increase in $N_{\rm PC}$ of $5-15\%$. For binary-exchange encounters, this mass dependence is present, but weaker.

\cite{2008MNRAS.386..553I} found a factor of 3 difference in the number of UCXBs formed via PCs and present at 11 Gyr in GCs that are dynamical twins but have different metallicity (see their Table 7, where the ratio is 2.8).
In that study, which  did not separate LMXBs as bright as extragalactic LMXBs from quiescent LMXBs that are detectable only in Milky Way GCs,  the importance of this channel was not emphasized as it produces a relatively small fraction of all LMXBs. However, this channel is significant if one only considers bright LMXBs for comparison to extragalactic studies.
The reason behind the factor of 3 difference was not identified, but re-analysis of the simulations showed that this may be due to the different number of giants in the stellar populations that are present in coeval GCs.
The number of RGs at 11 Gyr in the population of a whole cluster of ``standard'' (metal-rich) model, 
is $1.85\pm0.05$ times bigger than that of their ``metal-poor'' model.
At the same time, we found that the mass-segregation of RGs in the core may play an 
additional role: in the cores of the clusters, the enhancement of RGs was a factor of $2.1\pm0.1$, likely due to different turn-off masses.
This suggests the number density enhancement in the core of a younger red GC may be a factor of 2.6,
since mass segregation is
likely underestimated in these simulations that are not fully
dynamically self-consistent \citep{2009ApJ...707.1533F}. 

\cite{Kim12} also showed that red GCs may contain more high-luminosity LMXBs (those with $L_{\rm X}> 5\times 10^{38} {\rm \, erg \, s}^{-1}$ and thus most likely to have BH accretors) than blue GCs, with a ratio of $2.5^{+0.9}_{-1.1}$.
The formation of BH XRBs differs from NS XRBs -- BH XRBs are likely formed via sequences of multiple encounters occuring 
with seed BH-WD binaries that were formed in an initiating dynamical encounter  \citep{2010ApJ...717..948I}.
Since PCs with RGs provide at least half of all these seed BH-WD binaries, the above metallicity-dependent enrichment of RGs will also lead to the production of more BH-WD XRBs in red GCs; an exact prediction of this overproduction is beyond the scope of current dynamical codes.

\section{Enhancement of apperance probability}

\subsection{UCXBs}

The formation of an UCXB via PC between a NS and a RG depends not only on the probability of such a collision to occur, but also on the capacity of that formed binary to shrink sufficiently via gravitational wave radiation to start MT.
The latter, for each set of companion masses and post-collisional eccentricity and separation, can be determined using equations from \cite{1964PhRv..136.1224P}.
Here, we denote the maximum post-collisional separation in a post-collisional binary  
as a function of eccentricity that is capable of making an UCXB as $a_{\rm UCXB}(e)$: 
           only PCs where $a_{\rm PC}<a_{\rm UCXB}(e_{\rm PC})$ lead  to UCXB formation.

The post-collisional binary parameters $a_{\rm PC}$ and $e_{\rm PC}$  are 
functions of  the stellar masses involved in the PC,  
the radius of the  RG, $R_{\rm RG}$,  
the initial closest approach during the PC, 
$r_{\rm p}$, and $v_{\infty}$ (the latter parameter does not influence  outcomes  in GCs).
Hydrodynamical  studies  of  PCs in GCs between RGs and NSs for various $R_{\rm G}$  and  $r_{\rm P}$ 
have been  performed  by \cite{2006ApJ...640..441L};
the   parametrization  of  their  results, $(a_{\rm PC}; e_{\rm PC})=F(r_{\rm P}, R_{\rm G})$, was used as the
 underlying physics in population studies of LMXB formations in \cite{2008MNRAS.386..553I}.  

However,  \cite{2006ApJ...640..441L} only investigated two RG masses, 0.8  and 0.9  $M_\odot$. 
Thus, any mass dependence was not thoroughly explored, 
although their Fig.16 suggests that for larger $r_{\rm P}$ more massive RGs would form somewhat closer binaries.
Indeed, energy conservation implies that the formation of close binaries will be a function of the RG's 
envelope mass $M_{\rm env}$, as the latter determines how much energy has to be spend  on its ejection.
Neglecting the kinetic energy at infinity before and after the collision, a simplified estimate leads to $1/a_{\rm  PC}\propto M_{\rm env}$.

Note that RGs  located in blue  and red GCs of the same age 
would have different envelope masses  for the same core mass -- e.g. for a core of $0.3 M_\odot$, 
$M_{\rm env}$  differs by $50\%$. The resulting formed  binaries are expected to be more  compact  in  the  case of  a  red cluster.
Hence RGs in red GCs can form a binary (potentially an UCXB) in encounters that have a larger initial closest approach
$r_{\rm UCXB, red}$ than RGs in blue clusters. The ratio of these maximum closest approaches leading to an UCXB formation 
is $r_{\rm UCXB, red} / r_{\rm UCXB, blue} \sim {M_{\rm env, red}/M_{\rm env, blue}}$.
Hence, the ratio of encounter cross-sections resulting in UCXB formation in red and blue clusters is: 

\begin{equation}
\frac{\sigma_{\rm UCXB, red}}{\sigma_{\rm UCXB, blue}}\sim \frac{M_{\rm env, red}}{M_{\rm env, blue}} 
\left  ( \frac{M_{\rm RG, red}+M_{\rm NS}}{M_{\rm RG, blue}+M_{\rm NS}} \right ) \sim 1.3-1.6 \ . 
\end{equation}

\noindent Combined with \S~\ref{sec:encounter_rate}, this predicts that 
metal-rich GCs should contain a factor of $>2.7$ more UCXBs than coeval metal-poor GCs, 
in agreement with the results of the simulations. 
Note that since a PC does not lead to immediate UCXB formation, the relevant 
$M_{\rm env}$  is bigger than that of current RGs. However, 
in all the cases the appropriate $M_{\rm env}$  is bigger for metal-rich GCs compared to metal-poor ones.

We tested the expected dependence on the envelope mass by performing several PC simulations 
between a NS and RGs of different masses, $0.8~M_\odot$ and $1.1\ M_\odot$,
but the same $R_{\rm RG}$ and $M_{\rm core}$,
using the 3D SPH code {\tt StarCrash} \citep{Gaburov10,Lombardi11}. 
RGs'  stellar structures were first calculated using the 
{\tt STARS/ev} code and then relaxed in  {\tt StarCrash} 
(see \citealt{Glebbeek2008} and references therein).
The comparison of PC outcomes with the same $r_{\rm P}$ but different $M_{\rm env}$
confirmed that post-collisional binary separation is a function of $M_{\rm env}$ in the expected way -- smaller for larger $M_{\rm env}$,
although the dependence is a bit weaker than a strictly linear proportionality 
(we obtain ratios of post-collisional orbital separations being $\sim 10\%$ different from $M_{\rm env, red}/M_{\rm env, blue}$).

\subsection{LMXBs with RG donors}

\begin{figure}
\plotone{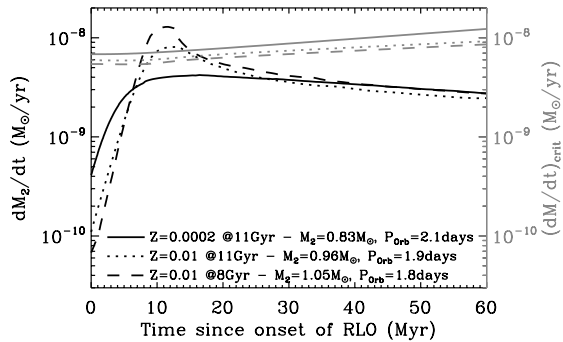}
\caption{MT rate (black lines) as a function of time passed from the onset of RLOF for three indicative MT sequences. In all case the onset of the RLOF happened when the donor star reached a radius of $\sim 3\,\rm R_{\odot}$, and the MT was assumed to be conservative up to the Eddington limit. For comparison, the critical MT below which the accretion disks become thermally unstable is plotted with grey color for each MT sequence. Only the higher metallicity systems undergo periods of persistent MT.
\label{rlo_MT}}
\end{figure}

Given $M_{\rm RG, red}$ and $M_{\rm RG, blue}$, the mass ratio of the dynamically formed LMXBs at 
the onset of the Roche-lobe overflow (RLOF) will be close to unity ($q=M_2/M_{\rm NS}\simeq 0.65-0.85$).
The MT rate that these binaries will drive depends strongly on the mass ratio of the binary.
Specifically, the higher the mass ratio, the higher the MT rate that the binary will drive.
Hence, LMXBs in red GC, which have higher mass ratios at the onset of the Roche-lobe overflow compared to those in blue GCs, are expected to drive overall higher MT rates.

To more properly study this effect, we used the {\tt MESA} code to calculate three indicative MT sequences between a giant donor star and a NS accretor, corresponding to a typical LMXB with a giant donor in an $11\,\rm Gyr$ old red GC, in an $8\,\rm Gyr$ old red GC, and in an $11\,\rm Gyr$ old blue GC. Figure~\ref{rlo_MT} shows these calculated MT rates as a function of time since the onset of RLOF. The RLOF always began when the donor star reached a radius of $\sim 3\,\rm R_{\odot}$, and the MT was assumed to be conservative up to the Eddington limit. For comparison, Figure~\ref{rlo_MT} also shows the critical MT below which the accretion disk becomes thermally unstable \citep{1999MNRAS.303..139D}. LMXBs with MT rates above the critical MT rate are believed to be persistent X-ray sources, while transient LMXBs have long-term average MT rates below this critical value. 

Figure~\ref{rlo_MT} shows that the MT rate in the two high-metallicity cases reaches peak values of 2.5--5 times higher than the low metallicity case, right after the onset of the RLO. During this initial phase, LMXBs with giant donors become persistent X-ray sources in red GCs, while they always remain transient sources in blue GCs. As a result, for the same total number of LMXBs with giant donors, we expect to be able to detect more bright X-ray sources in red GCs than blue GCs. 

\section{Discussion}

In this {\it Letter} we attempted to explain the observed ratio of bright LMXBs in extragalactic GCs of different metallicities.
We identified that the RG population capable of forming close binaries is both more abundant  (as a fraction among all of the stars) and more massive in metal-rich clusters  than metal-poor  clusters.
We propose that these properties lead to strong consequences on the number of bright LMXBs that can be formed and observed in otherwise dynamically similar clusters (total GC mass, star's number density and velocity dispersion).
First, more close binary systems are formed, both via physical collision or exchange encounters --
the increase is a factor of 1.6--2.6 (their fraction among all the stars).
Secondly, a higher fraction of formed NS-WD will be able to start MT as bright UCXBs, as their post-collisional configurations are more compact. 
Finally, LMXBs with high-metallicity giant donors drive higher MT rates and can appear as persistent systems, while low-metallicity LMXBs with giant donors are transient. 
The combination of these factors is capable of producing the observed overabundance of bright LMXBs in red GCs compared to blue GCs, although a full population synthesis study that includes proper physics
of both collisions and MT with RGs is necessary to quantitatively account for the combined effect.

\acknowledgments
NI and GRS acknowledge support by  NSERC Discovery Grants. 
TF acknowledges support from the CfA and the ITC prize fellowship programs.
DWK and GF acknowledge support from NASA contract NAS8-39073 (CXC).
JLAN acknowledges support from CONACyT.
AJ is supported by the Chilean Ministry for the Economy, Development,
and Tourism's Programa Iniciativa Cient\'{i}fica Milenio through grant
P07-021-F, awarded to The Milky Way Millennium Nucleus.
This material is based upon work supported in part by the National Science Foundation under Grant No. 1066293 and the hospitality of the Aspen Center for Physics. 

%\bibliography{gc}

\begin{thebibliography}{34}
\expandafter\ifx\csname natexlab\endcsname\relax\def\natexlab#1{#1}\fi

\bibitem[{{Bellazzini} {et~al.}(1995){Bellazzini}, {Pasquali}, {Federici},
  {Ferraro}, \& {Pecci}}]{1995ApJ...439..687B}
{Bellazzini}, M., {Pasquali}, A., {Federici}, L., {Ferraro}, F.~R., \& {Pecci},
  F.~F. 1995, \apj, 439, 687

\bibitem[{{Brodie} \& {Strader}(2006)}]{2006ARA&A..44..193B}
{Brodie}, J.~P. \& {Strader}, J. 2006, \araa, 44, 193

\bibitem[{{Clark}(1975)}]{1975ApJ...199L.143C}
{Clark}, G.~W. 1975, \apjl, 199, L143

\bibitem[{{Dubus} {et~al.}(1999){Dubus}, {Lasota}, {Hameury}, \&
  {Charles}}]{1999MNRAS.303..139D}
{Dubus}, G., {Lasota}, J.-P., {Hameury}, J.-M., \& {Charles}, P. 1999, \mnras,
  303, 139

\bibitem[{{Fan} \& {de Grijs}(2012)}]{2012arXiv1205.4310F}
{Fan}, Z. \& {de Grijs}, R. 2012, ArXiv e-prints

\bibitem[{{Fragos} {et~al.}(2008){Fragos}, {Kalogera}, {Belczynski},
  {Fabbiano}, {Kim}, {Brassington}, {Angelini}, {Davies}, {Gallagher}, {King},
  {Pellegrini}, {Trinchieri}, {Zepf}, {Kundu}, \&
  {Zezas}}]{2008ApJ...683..346F}
{Fragos}, T., {Kalogera}, V., {Belczynski}, K., {Fabbiano}, G., {Kim}, D.-W.,
  {Brassington}, N.~J., {Angelini}, L., {Davies}, R.~L., {Gallagher}, J.~S.,
  {King}, A.~R., {Pellegrini}, S., {Trinchieri}, G., {Zepf}, S.~E., {Kundu},
  A., \& {Zezas}, A. 2008, \apj, 683, 346

\bibitem[{{Fragos} {et~al.}(2009){Fragos}, {Kalogera}, {Willems}, {Belczynski},
  {Fabbiano}, {Brassington}, {Kim}, {Angelini}, {Davies}, {Gallagher}, {King},
  {Pellegrini}, {Trinchieri}, {Zepf}, \& {Zezas}}]{2009ApJ...702L.143F}
{Fragos}, T., {Kalogera}, V., {Willems}, B., {Belczynski}, K., {Fabbiano}, G.,
  {Brassington}, N.~J., {Kim}, D.-W., {Angelini}, L., {Davies}, R.~L.,
  {Gallagher}, J.~S., {King}, A.~R., {Pellegrini}, S., {Trinchieri}, G.,
  {Zepf}, S.~E., \& {Zezas}, A. 2009, \apjl, 702, L143

\bibitem[{{Fregeau} {et~al.}(2009){Fregeau}, {Ivanova}, \&
  {Rasio}}]{2009ApJ...707.1533F}
{Fregeau}, J.~M., {Ivanova}, N., \& {Rasio}, F.~A. 2009, \apj, 707, 1533

\bibitem[{{Gaburov} {et~al.}(2010){Gaburov}, {Lombardi}, \& {Portegies
  Zwart}}]{Gaburov10}
{Gaburov}, E., {Lombardi}, Jr., J.~C., \& {Portegies Zwart}, S. 2010, \mnras,
  402, 105

\bibitem[{{Glebbeek} {et~al.}(2008){Glebbeek}, {Pols}, \&
  {Hurley}}]{Glebbeek2008}
{Glebbeek}, E., {Pols}, O.~R., \& {Hurley}, J.~R. 2008, \aap, 488, 1007

\bibitem[{{Grindlay}(1993)}]{1993ASPC...48..156G}
{Grindlay}, J.~E. 1993, in Astronomical Society of the Pacific Conference
  Series, Vol.~48, The Globular Cluster-Galaxy Connection, ed. G.~H. {Smith} \&
  J.~P. {Brodie}, 156

\bibitem[{{Ivanova}(2006)}]{2006ApJ...636..979I}
{Ivanova}, N. 2006, \apj, 636, 979

\bibitem[{{Ivanova} {et~al.}(2010){Ivanova}, {Chaichenets}, {Fregeau},
  {Heinke}, {Lombardi}, \& {Woods}}]{2010ApJ...717..948I}
{Ivanova}, N., {Chaichenets}, S., {Fregeau}, J., {Heinke}, C.~O., {Lombardi},
  Jr., J.~C., \& {Woods}, T.~E. 2010, \apj, 717, 948

\bibitem[{{Ivanova} {et~al.}(2008){Ivanova}, {Heinke}, {Rasio}, {Belczynski},
  \& {Fregeau}}]{2008MNRAS.386..553I}
{Ivanova}, N., {Heinke}, C.~O., {Rasio}, F.~A., {Belczynski}, K., \& {Fregeau},
  J.~M. 2008, \mnras, 386, 553

\bibitem[{{Ivanova} {et~al.}(2005){Ivanova}, {Rasio}, {Lombardi}, {Dooley}, \&
  {Proulx}}]{2005ApJ...621L.109I}
{Ivanova}, N., {Rasio}, F.~A., {Lombardi}, Jr., J.~C., {Dooley}, K.~L., \&
  {Proulx}, Z.~F. 2005, \apjl, 621, L109

\bibitem[{{Jord{\'a}n} {et~al.}(2004){Jord{\'a}n}, {C{\^o}t{\'e}}, {Ferrarese},
  {Blakeslee}, {Mei}, {Merritt}, {Milosavljevi{\'c}}, {Peng}, {Tonry}, \&
  {West}}]{2004ApJ...613..279J}
{Jord{\'a}n}, A., {C{\^o}t{\'e}}, P., {Ferrarese}, L., {Blakeslee}, J.~P.,
  {Mei}, S., {Merritt}, D., {Milosavljevi{\'c}}, M., {Peng}, E.~W., {Tonry},
  J.~L., \& {West}, M.~J. 2004, \apj, 613, 279

\bibitem[{{Kalogera} {et~al.}(2004){Kalogera}, {King}, \&
  {Rasio}}]{2004ApJ...601L.171K}
{Kalogera}, V., {King}, A.~R., \& {Rasio}, F.~A. 2004, \apjl, 601, L171

\bibitem[{{Kim} {et~al.}(2006){Kim}, {Kim}, {Fabbiano}, {Lee}, {Park},
  {Geisler}, \& {Dirsch}}]{2006ApJ...647..276K}
{Kim}, E., {Kim}, D.-W., {Fabbiano}, G., {Lee}, M.~G., {Park}, H.~S.,
  {Geisler}, D., \& {Dirsch}, B. 2006, \apj, 647, 276

\bibitem[{Kim} {et~al.}(2012)]{Kim12}
{Kim}, D.-W., et al., 2012, \apj, submitted

\bibitem[{{Kroupa}(2001)}]{2001MNRAS.322..231K}
{Kroupa}, P. 2001, \mnras, 322, 231

\bibitem[{{Kundu} {et~al.}(2002){Kundu}, {Maccarone}, \&
  {Zepf}}]{2002ApJ...574L...5K}
{Kundu}, A., {Maccarone}, T.~J., \& {Zepf}, S.~E. 2002, \apjl, 574, L5

\bibitem[{{Kundu} {et~al.}(2007){Kundu}, {Maccarone}, \&
  {Zepf}}]{2007ApJ...662..525K}
---. 2007, \apj, 662, 525

\bibitem[{{Lombardi} {et~al.}(2011){Lombardi}, {Holtzman}, {Dooley}, {Gearity},
  {Kalogera}, \& {Rasio}}]{Lombardi11}
{Lombardi}, Jr., J.~C., {Holtzman}, W., {Dooley}, K.~L., {Gearity}, K.,
  {Kalogera}, V., \& {Rasio}, F.~A. 2011, \apj, 737, 49

\bibitem[{{Lombardi} {et~al.}(2006){Lombardi}, {Proulx}, {Dooley}, {Theriault},
  {Ivanova}, \& {Rasio}}]{2006ApJ...640..441L}
{Lombardi}, Jr., J.~C., {Proulx}, Z.~F., {Dooley}, K.~L., {Theriault}, E.~M.,
  {Ivanova}, N., \& {Rasio}, F.~A. 2006, \apj, 640, 441

\bibitem[{{Maccarone} {et~al.}(2004){Maccarone}, {Kundu}, \&
  {Zepf}}]{2004ApJ...606..430M}
{Maccarone}, T.~J., {Kundu}, A., \& {Zepf}, S.~E. 2004, \apj, 606, 430

\bibitem[{{Paxton} {et~al.}(2011){Paxton}, {Bildsten}, {Dotter}, {Herwig},
  {Lesaffre}, \& {Timmes}}]{2011ApJS..192....3P}
{Paxton}, B., {Bildsten}, L., {Dotter}, A., {Herwig}, F., {Lesaffre}, P., \&
  {Timmes}, F. 2011, \apjs, 192, 3

\bibitem[{{Peacock} {et~al.}(2010){Peacock}, {Maccarone}, {Kundu}, \&
  {Zepf}}]{2010MNRAS.407.2611P}
{Peacock}, M.~B., {Maccarone}, T.~J., {Kundu}, A., \& {Zepf}, S.~E. 2010,
  \mnras, 407, 2611

\bibitem[{{Peters}(1964)}]{1964PhRv..136.1224P}
{Peters}, P.~C. 1964, Physical Review, 136, 1224

\bibitem[{{Revnivtsev} {et~al.}(2011){Revnivtsev}, {Postnov}, {Kuranov}, \&
  {Ritter}}]{2011A&A...526A..94R}
{Revnivtsev}, M., {Postnov}, K., {Kuranov}, A., \& {Ritter}, H. 2011, \aap,
  526, A94

\bibitem[{{Sidoli} {et~al.}(2001){Sidoli}, {Parmar}, {Oosterbroek}, {Stella},
  {Verbunt}, {Masetti}, \& {Dal Fiume}}]{2001A&A...368..451S}
{Sidoli}, L., {Parmar}, A.~N., {Oosterbroek}, T., {Stella}, L., {Verbunt}, F.,
  {Masetti}, N., \& {Dal Fiume}, D. 2001, \aap, 368, 451

\bibitem[{{Sivakoff} {et~al.}(2007){Sivakoff}, {Jord{\'a}n}, {Sarazin},
  {Blakeslee}, {C{\^o}t{\'e}}, {Ferrarese}, {Juett}, {Mei}, \&
  {Peng}}]{2007ApJ...660.1246S}
{Sivakoff}, G.~R., {Jord{\'a}n}, A., {Sarazin}, C.~L., {Blakeslee}, J.~P.,
  {C{\^o}t{\'e}}, P., {Ferrarese}, L., {Juett}, A.~M., {Mei}, S., \& {Peng},
  E.~W. 2007, \apj, 660, 1246

\bibitem[{{Timmes} {et~al.}(1996){Timmes}, {Woosley}, \&
  {Weaver}}]{1996ApJ...457..834T}
{Timmes}, F.~X., {Woosley}, S.~E., \& {Weaver}, T.~A. 1996, \apj, 457, 834

\bibitem[{{Trudolyubov} \& {Priedhorsky}(2004)}]{2004ApJ...616..821T}
{Trudolyubov}, S. \& {Priedhorsky}, W. 2004, \apj, 616, 821

\bibitem[{{Verbunt} \& {Lewin}(2006)}]{2006csxs.book..341V}
{Verbunt}, F. \& {Lewin}, W.~H.~G. {Globular cluster X-ray sources}, ed.
  W.~H.~G. {Lewin} \& M.~{van der Klis}, 341--379

\bibitem[{{Woodley} {et~al.}(2010){Woodley}, {Harris}, {Puzia}, {G{\'o}mez},
  {Harris}, \& {Geisler}}]{2010ApJ...708.1335W}
{Woodley}, K.~A., {Harris}, W.~E., {Puzia}, T.~H., {G{\'o}mez}, M., {Harris},
  G.~L.~H., \& {Geisler}, D. 2010, \apj, 708, 1335

\end{thebibliography}

\end{document}